\documentclass[a4paper,12pt]{article}
\usepackage{graphicx}
\usepackage[]{amssymb}
\begin{document}

\title{Continuation of periodic orbits in two-planet resonant systems}
\author{G. Voyatzis, T. Kotoulas and  J.D. Hadjidemetriou
\\University of Thessaloniki, Department of Physics\\
Section of Astrophysics, Astronomy and Mechanics\\
GR-541 24 Thessaloniki, Greece\\
e-mail address: voyatzis@auth.gr}
\date{}

\maketitle

\begin{abstract}
The continuation of resonant periodic orbits from the restricted to the general three body problem is studied in a systematic way. Starting from the Keplerian unperturbed system we obtain the resonant families of the circular restricted problem. Then we find all the families of the resonant elliptic restricted three body problem which bifurcate from the circular model. All these families are continued to the general three body problem, and in this way we can obtain a global picture of all the families of periodic orbits of a two-planet resonant system. We consider planar motion only. We show that the continuation follows a scheme proposed by Bozis and Hadjidemetriou (1976) for symmetric orbits. Our study includes also asymmetric periodic orbits, which exist in cases of external resonances. The families formed by passing from the restricted to the general problem are continued within the framework of the general problem by varying the planetary mass ratio $\rho$. We obtain bifurcations which are caused either due to collisions of the families in the space of initial conditions or due to the vanishing of bifurcation points. Our study refers to the whole range of planetary mass ratio values ($\rho \in (0,\infty)$) and, therefore we include the passage from external to internal resonances. In the present work, our numerical study includes the case of the 2/1 and 1/2 resonance. The same method can be used to study all other planetary resonances.
\end{abstract}


\section{Introduction} 
A good model to study the motion  of three celestial bodies considered as point masses, e.g. a triple stellar system or a planetary system with two planets, is the famous {\em three-body problem} (TBP), whose study goes back to Poincar\'e. In the present study we consider the case where only one of the three bodies is the more massive one, and the other two bodies have much smaller masses. This model is useful in the study of the motion of small bodies in a planetary system (for example asteroids and comets in our Solar System) or planetary systems with two planets.

The simplest model is the {\em circular restricted TBP}. Although
much work has been carried out for this model, (see e.g. Bruno,
1994; H\'enon, 1997), new interesting results continue to appear in
the literature (Maciejewski and Rybicki 2004; Papadakis and Goudas,
2006; Bruno and Varin 2006,2007). In this model we consider two
bodies with non zero mass, called {\em primaries}, for example the
Sun and Jupiter, moving in circular orbits around their common
center of mass, and a third body with negligible mass, for example
an asteroid, moves under the gravitational attraction of the two
primaries. A more realistic model is the {\em elliptic restricted
TBP}, where the two primaries move in elliptic orbits, with a finite
eccentricity. However, an important aspect of the dynamics is
missing in these two models. It is the gravitational interaction
between the small body and the two primaries, which is not taken
into account in the restricted models. When we introduce this
gravitational interaction, we have a more realistic model, the {\em
general TBP}. Within this framework we study a system consisting of
a Sun and two small bodies, which we call {\em planets}.

In the study of a dynamical system, the topology of its phase space
plays a crucial role. The topology is determined by the position and
the stability properties of the periodic orbits, or equivalently, of
the fixed points of the Poincar\'e map on a surface of section. This
makes clear the importance of the knowledge of the families of
periodic orbits in a dynamical system. Particularly, in a planetary
system, many families of periodic orbits are associated with
resonances, which are mean motion resonances between the two
planets.  Since in our study of planetary systems only one body is
the more massive one (the sun), a good method is to start from the
simplest model, which is the circular restricted problem and find
all the basic families of periodic orbits. Then we extend the model
to the elliptic restricted model, and find all the families of
resonant periodic orbits that bifurcate from the circular to the
elliptic model. Finally, we give mass to the massless body and
continue all these families to the model of the general problem.
This is the method that we shall use in the present study to
describe the topology of the phase space. The existence of periodic
orbits of the planetary type in the general TBP, as a continuation
from the restricted problem, has been studied by Hadjidemetriou
(1975, 1976) and recently this method found a fruitful field of
applicability in the dynamics of resonant extrasolar systems (e.g.
Rivera and Lissauer,2001; Hadjidemetriou, 2002;  Ji et al, 2003;
Haghighipour et al., 2003; Ferraz-Mello et al., 2003, 2005;
Psychoyos and Hadjidemetriou, 2005; Voyatzis and Hadjidemetriou,
2005, 2006; Voyatzis, 2008).  An alternative method for determining
resonant periodic orbits is the computation of stationary solutions
of an averaged model (Beauge et al. 2003; Michtchenko at al. 2006).

It is interesting to mention at this point that the situation is not the same in all resonances. It may happen that in some resonances there are not new families in the elliptic model, bifurcating from the circular model, or may exist several resonant families bifurcating from the circular model. This has, evidently, important consequences on the topology of the phase space and differentiates the behavior of the model in different resonances (see Tsiganis et al. 2002a, 2002b). Although in this study we present numerical computations of the 2/1 (or 1/2) resonance, our approach and results have a more general applicability.

In the next section we discuss briefly known aspects on the families of periodic orbits in the circular unperturbed and restricted problem. However, these issues are fundamental for continuing our study in the elliptic restricted and in the general problem. In section \ref{ERTBP} we present the continuation of resonant periodic orbits from the circular to the elliptic model and then, in section \ref{SectionCon}, we consider the continuation in the general problem. In section \ref{GTBPbif} we study the bifurcation of families of periodic orbits within the framework of the general problem and finally, in section \ref{CD}, we discuss the generality of our results and conclude.

\section{The circular restricted problem}

Consider a body $S$ (Sun) with mass $m_0$ and a second body $J$ (Jupiter) with mass $m_1$, which describe circular orbits around their common center of mass. We define a rotating frame of reference $xOy$, whose $x$-axis is the line $SJ$, the origin is at their center of mass and the $xy$ plane is the orbital plane of the circular motion of the these two bodies. The circular restricted problem describes the motion of a massless body $A$ in the rotating frame, which moves under the gravitational attraction of $S$ and $J$. In our computations we consider the normalization of units $m_0+m_1=1$,   $G=1$ and $n'=1$, where $G$ is the gravity constant and $n'$ the mean motion of $J$.

%%Fig. 1
\begin{figure}[tb]
\centering
\includegraphics[height=6cm]{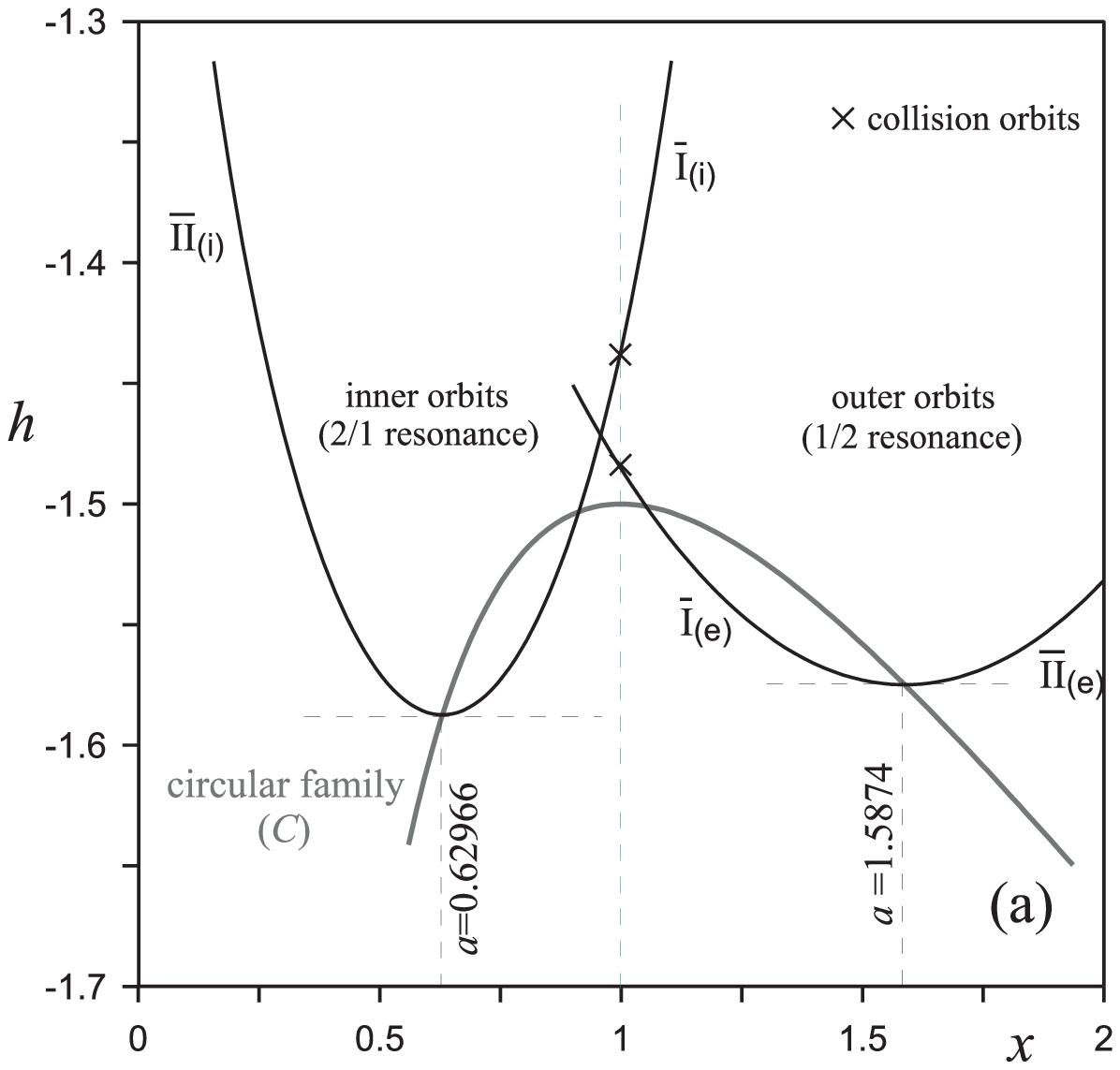} \hspace{1cm}
\includegraphics[height=5.9cm]{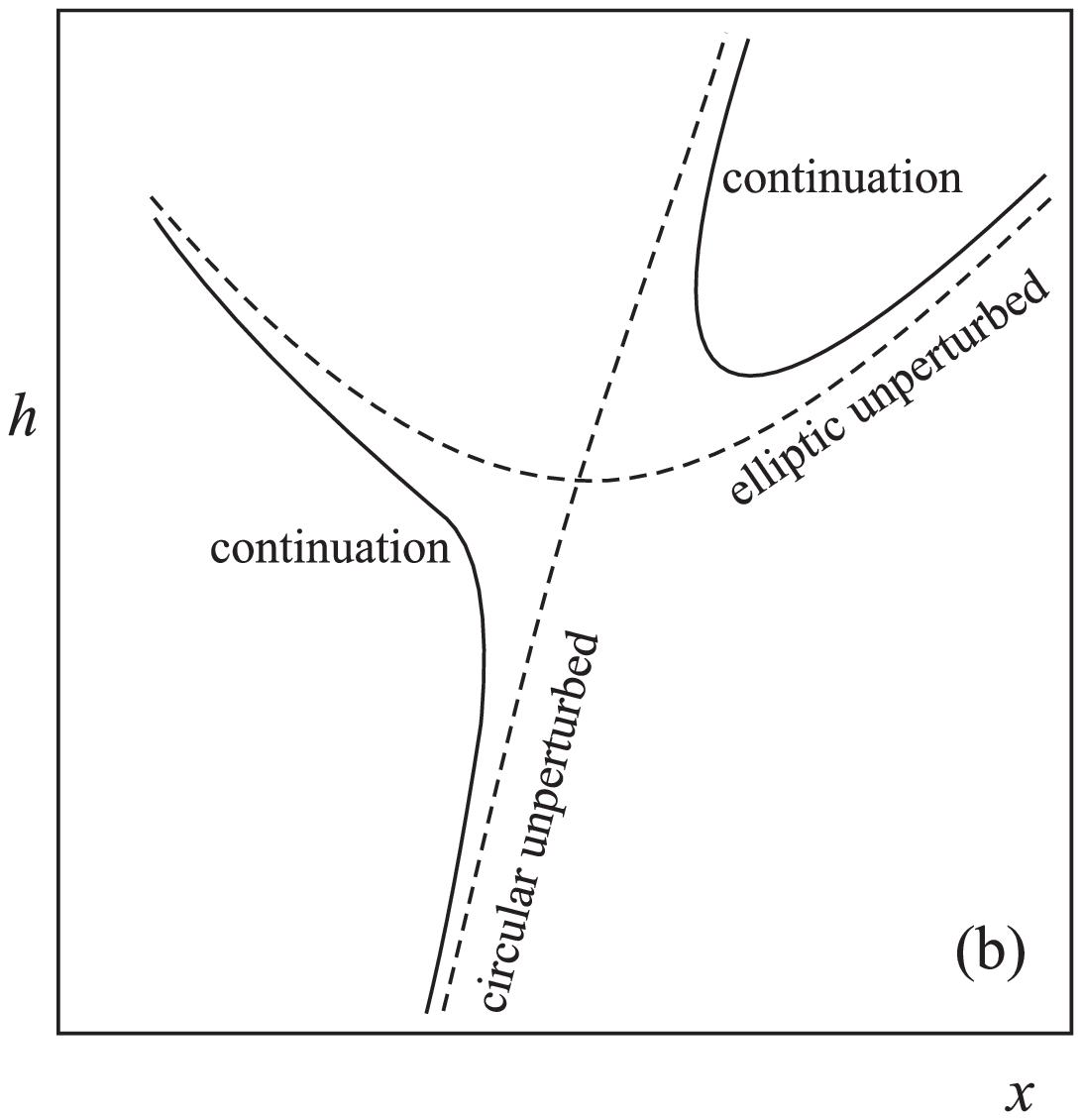}
\caption{(a) The unperturbed case where Jupiter mass is zero ($m_1=0$). The family of circular periodic orbits and the the two resonant families at the 2/1 and 1/2 resonance that bifurcate from the circular family. (b) The bifurcation to the two 2/1 resonant elliptic families (inner resonance), when the mass of Jupiter is non zero. The same topology appears at the bifurcation point at the 1/2 resonance (outer resonance).}
\label{UnFam}
\end{figure}

\subsection{The unperturbed case ($m_1=0$)}
If $J$ is assumed to be a massless body too, then in the inertial
frame the body $A$ can move in a Keplerian ellipsis of semimajor
axis $a$, eccentricity $e$ and mean motion $n\sim a^{-3/2}$.
Circular orbits ($e=0$) with radious $r=r_0=a$ are also included.
The above mentioned orbits are periodic in the inertial frame.

In the rotating frame, the Hamiltonian  that describes the unperturbed motion of $A$, in polar coordinates, $r$, $\phi$, is
\begin{equation}
H_0=\frac{p_r^2}{2}+\frac{p_\phi^2}{2r^2}-n' p_\phi-\frac{G m_0}{r}.
\end{equation}
The momenta are $p_r=\dot r$ and $p_\phi=r^2(\dot
\phi+n')$=constant. In the rotating frame periodic orbits also exist
and are of two kinds:

\noindent{\em Circular orbits}. In the rotating frame there exist circular orbits of the body $A$ with an arbitrary radius $r=r_0$ and are evidently symmetric with respect to the rotating $x$-axis. Consequently, a {\it family of circular periodic orbits} exists along which the radius $r_0=a$ or the frequency $n=(G m_0)^{1/2} a^{-3/2}$ varies. This family is represented by a smooth curve in the space $x_0-h$, where $x_0$ is the initial condition of the orbit, which is $x_0=r_0$ or, equivalently, $x_0=a$, and $h$ denotes the value of the energy integral $H_0$. The family is given by the equation
\begin{equation}
-\frac{G m_0}{2x_0}-n'\sqrt{G m_0 x_0}=h,
\label{UPFAM}
\end{equation}

\noindent{\em Elliptic orbits}. An elliptic orbit of the small body $A$ in the inertial frame is periodic in the
rotating frame only if it is resonant, i.e. $n/n'=p/q$, with $p,q$ integers. If $r_0=a_1$ is the semimajor axis (radious) of Jupiter then a $n/n'=p/q$ resonance corresponds for the massless body A to a semimajor axis $a=a_{p/q}=(n/n')^{-2/3}$ and is periodic for any eccentricity $e$. Thus  a {\em family of resonant elliptic periodic orbits} exists, with the eccentricity as a parameter along the family. The ratio $n/n'$ is almost constant along the family. There is however another parameter, defining the {\em orientation} of the elliptic orbit, which is the angle $\omega$ of the line of apsides with a fixed direction. If this fixed direction is defined by a conjunction of the three bodies then an elliptic periodic orbit is symmetric when $\omega=0$ or $\omega=\pi$.

In the space $x_0-h$, an elliptic family is represented by a smooth curve, given by the equation (\ref{UPFAM}) where now $x_0=a_{p/q}(1-e)$ is the
pericenter distance. Note that this presentation is not unique: a point on the elliptic family represents {\it all} the elliptic resonant orbits with the same eccentricity, but arbitrary orientation $\omega$. An elliptic periodic orbit in the rotating frame is also periodic in the inertial frame.
Along the circular family the value of the semimajor axis varies and, consequently, the ratio $n/n'$ varies and passes through resonant values $a=a_{p/q}$. It is at these points that we have a bifurcation to an elliptic family.

The circular family and the 2/1 and 1/2 resonant families of periodic orbits of the unperturbed problem are presented in Fig.\ref{UnFam}a. In the normalization we are using, the semi major axes of the 2/1 family are all $a<1$, called {\it internal resonance} and the semimajor axes of the 1/2 resonance are all $a>1$, called {\it external resonance}. The bifurcation points are $r_0=a_{2/1}=0.6297$ and $r_0=a_{1/2}=1.5874$, respectively. At the bifurcation points, the tangent to the above resonant elliptic families is parallel to the $x$-axis.

Each family of the internal and the external resonance, is divided into two parts by the corresponding bifurcation point. One part corresponds to position of the small body at perihelion and the other at aphelion. In particular, for the internal family, the part $x<0.6297$ corresponds to perihelion (family $\bar{II}_i$) and the part $x>0.6297$ corresponds to aphelion (family $\bar{I}_i$). For the external family, the part $x<1.5874$ corresponds to perihelion (family $\bar{I}_e$) and the part $x>1.5874$ corresponds to aphelion (family $\bar{II}_e$). Note that on the aphelion part of the inner family and on the perihelion part of the outer family, there is a collision orbit with Jupiter. Evidently, all the circular and the elliptic orbits are stable, as they are Keplerian orbits.

\subsection{The perturbed case: Non zero mass of Jupiter} \label{CRTBP}

Let us now assume that the mass of Jupiter is non zero i.e.
$m_1=\mu\neq 0$ and $m_0=1-\mu$. The resonant elliptic families at
the 2/1 and 1/2 resonance are continued to $\mu>0$, but a gap
appears at the bifurcation point of the unperturbed families . The
topology at the bifurcation point of the resonant families and the
gap that appears, is shown in Figure \ref{UPFAM}b. There are two
families of elliptic orbits in each resonance, one corresponding to
the case where the small body is at perihelion and the other to the
case where the small body is at aphelion.  All these orbits are {\it
symmetric} periodic orbits, i.e. out of the infinite set of all
symmetric and asymmetric orbits of the unperturbed problem, only two
orbits survive for $\mu>0$, both symmetric. One of them is stable
and the other unstable, as a consequence of the Poincar\'e-Birkhoff
fixed point theorem, but along the family the stability may change.

\begin{figure}[tb]
\centering
\includegraphics[width=7cm]{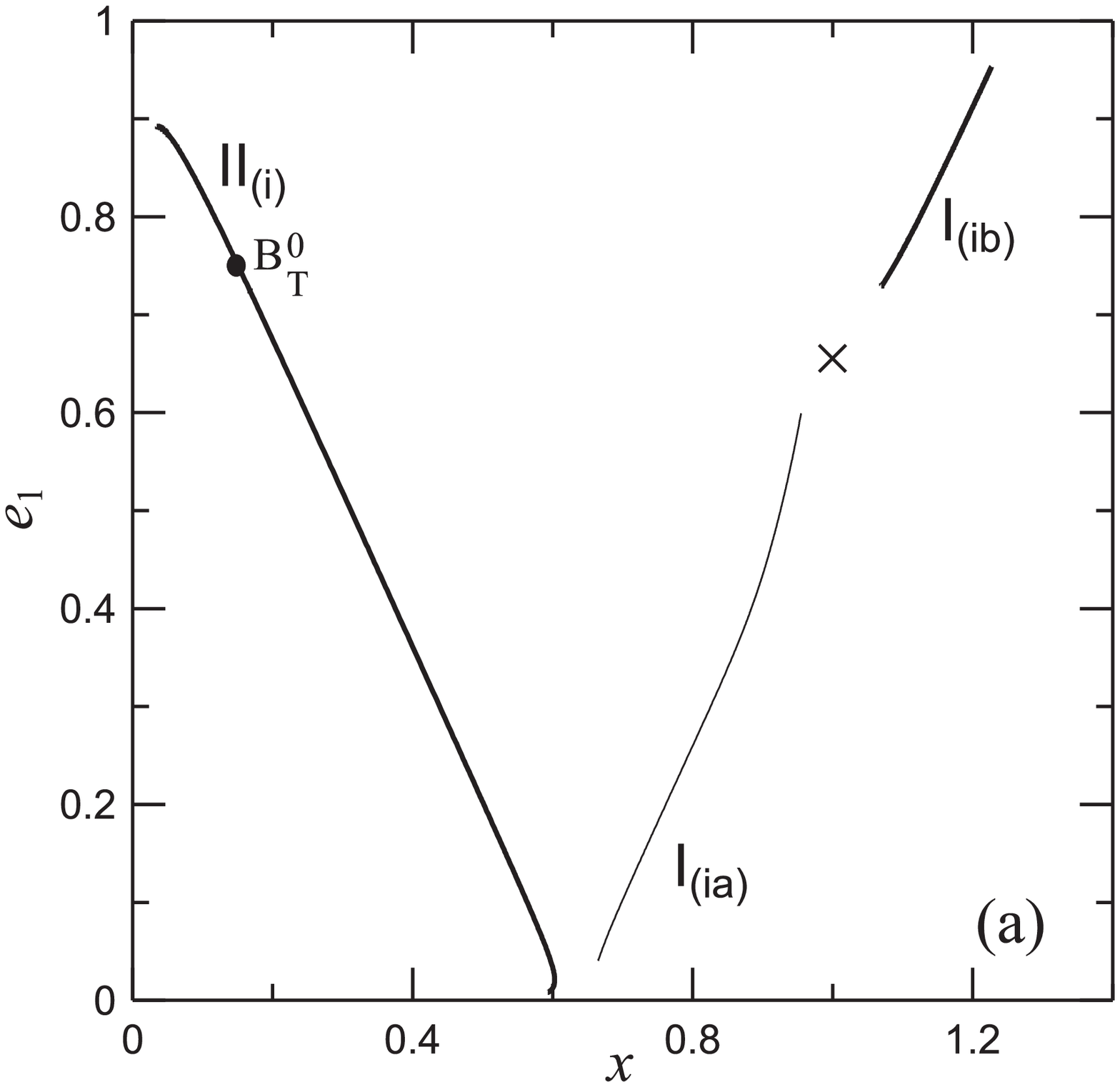} \hspace{1cm}
\includegraphics[width=7cm]{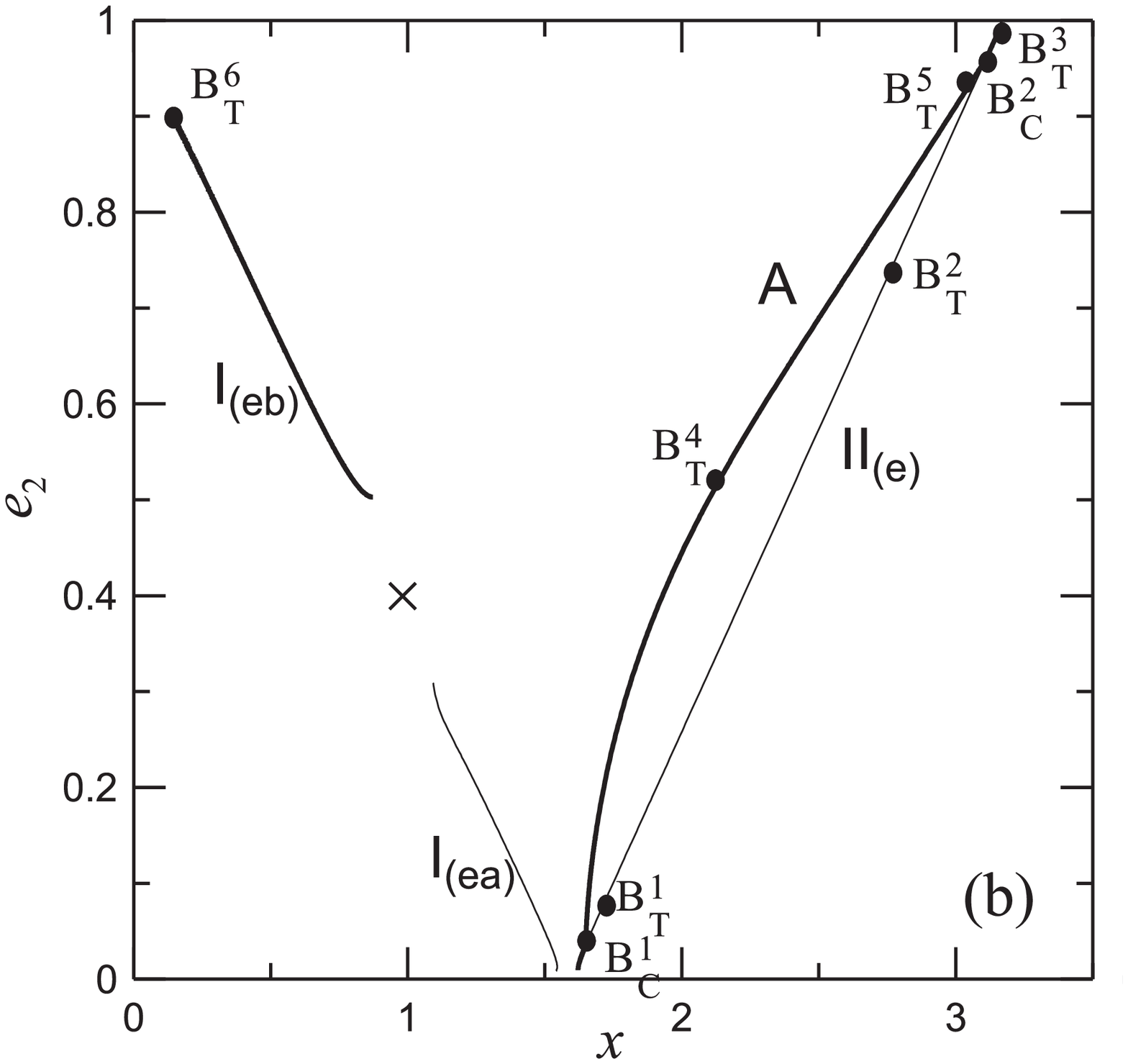}
\caption{Resonant families of elliptic periodic orbits of the
circular restricted problem. a) the case of internal 2/1 resonance.
b) the case of external 1/2 resonance. Thin or bold curves indicate
unstable or stable orbits, respectively. The symbol ''$\times$''
indicates a region of close encounters between Jupiter and the small
body. Note that we denote with $e_1$ the eccentricity of the small
body in the inner orbits and with $e_2$ the eccentricity of the
small body in the outer orbits.} \label{FCRPO}
\end{figure}

In Figure \ref{FCRPO}a we present the families of periodic orbits
for the 2/1 {\it internal} resonance, for $\mu=0.001$. The gap at
the bifurcation point that we showed in Figure \ref{UPFAM}b is
presented in this Figure by a small gap at the point $e_1\approx 0$
($x\approx 0.6297$).  The families are presented in a plane where
the horizontal axis indicates the $x$-coordinate of the massless
body in the rotating frame and the vertical axis corresponds to the
osculating eccentricity of the nearly Keplerian orbit of the
massless body. Correspondingly to the unperturbed case, there are
two families, family $I_{(i)}$, where the small body is at aphelion
and family $II_{(i)}$, where the small body is at perihelion. Family
$I_{(i)}$ consists of two parts: $I_{(ia)}$ and $I_{(ib)}$ separated
by a collision (see Figure \ref{UPFAM}a). The first part consists of
unstable orbits (thin curve) while the second part consists of
stable orbits (thick curve). The orbits of the family $II_{(i)}$ are
all stable.

In the case of the {\it external} 1/2 resonance, the corresponding
families are shown in Figure \ref{FCRPO}b. These families arise from
the 1/2 elliptic family of Figure \ref{UPFAM}a when the last one is
continued to $\mu>$0. As in the internal resonance 2/1, a small gap
appears at the point $e_2=0$, $x=1.5874$. There are two families,
family $I_{(e)}$, where the small body is at perihelion and family
$II_{(e)}$, where the small body is at aphelion. As in the internal
resonance 2/1, the family $I_{(e)}$ consists of an unstable part
($I_{(ea)}$) and a stable part ($I_{(eb)}$), which are separated by
a collision (see Figure \ref{UPFAM}a). The family $II_{(e)}$ starts
and ends with stable orbits, but along the family the stability
changes and an unstable part exists between the critical points
$B_C^1$ and $B_C^2$. From each one of these critical points there
bifurcates a family of {\it asymmetric} periodic orbits. It turns
out that these two asymmetric families coincide to a single
asymmetric family $A$, which starts from the point $B_C^1$ and ends
to the point $B_C^2$. We have found that this family is whole stable
for $\mu<5.2 10^{-3}$. We remark that the asymmetric family exists
only for the external resonances of the form $1/q$, called
asymmetric resonances (Beauge, 1994; Voyatzis et al., 2005).

\section{The elliptic restricted model} \label{ERTBP}
Along the resonant elliptic families of the circular restricted problem the period varies.
In the unperturbed case ($\mu=0$) the period is exactly equal to $2\pi$ along the family, for the normalization we are using. When $\mu>0$, the period along the elliptic resonant families varies, but is close to the value $2\pi$. If it happens that for a  particular orbit of the family the period is {\it exactly} equal to $T_0=2\pi$, this point is a bifurcation point for a family of periodic orbits of the elliptic restricted TBP, along which the eccentricity of the second primary (Jupiter) varies. The same is true in the general case where the period of a periodic orbit on a family of the circular restricted problem is a multiple of $\pi$ (Broucke, 1969a,b).

Concerning the stability, we remark that the monodromy matrix of the
variational equations has two pairs of eigenvalues
($\lambda_1$,$\lambda_2$) and ($\lambda_3$, $\lambda_4$). In the
circular model one pair, ($\lambda_1,\lambda_2$) is the unit pair,
$\lambda_1=\lambda_2=1$, because of the existence of the energy
integral. The other pair, ($\lambda_3,\lambda_4$), may lie on the
unit circle in the complex plane ($\lambda_{3,4}=e^{\pm i\phi})$,
corresponding to stability, or may be on the real axis
($\lambda_3=1/\lambda_4\in R$) corresponding to instability. In the
elliptic model there is no energy integral, so we have four
different cases: (i) {\em stable} orbits when all eigenvalues are on
the unit circle  (ii) {\em simply unstable} orbits when one pair of
eigenvalues is on the unit circle and one on the real axis (iii)
{\em doubly unstable} orbits when both pairs of eigenvalues are on
the real axis and (iv) {\it complex instability}, where all
eigenvalues are outside the unit circle, not on the real axis,
arranged in reciprocal pairs and complex conjugate pairs
$\lambda_{1,2}=Re^{\pm i\phi}$, $\lambda_{3,4}=R^{-1}e^{\pm i\phi}$
(see Broucke 1969a,b). We also can define the stability indices
$b_1=\lambda_1+\lambda_2$ and $b_2=\lambda_3+\lambda_4$ (see
Hadjidemetriou, 2006). A periodic orbit is stable if $$|b_i|<2,
\;\;\forall i=1,2.$$ If one of the indices $b_i$ does not satisfy
the stability condition, the periodic orbit is simply unstable. If
both indices do not satisfy the stability condition the periodic
orbit is doubly unstable. Complex instability is not present in the
particular model.

In the case of 2/1 and 1/2 resonant families, the critical points, i.e. periodic orbits with period exactly equal to $2\pi$, are indicated on the families $II_{(i)}$ and $II_{(e)}$ in Figure \ref{FCRPO} by the points $B_T^i, i=0,1,2,\dots 6$. Note that for the internal resonance 2/1, there is only one bifurcation point, $B_T^0$, on the stable family $II_{(i)}$. For the external resonance 1/2 there are five bifurcation points. Three of them, $B_T^1$, $B_T^2$ and $B_T^3$ belong to the symmetric family $II_{(e)}$ and two more points, $B_T^4$, $B_T^5$ belong to the asymmetric family $A$. There are no bifurcation points on the families $I_{(i)}$ while on the family $I_{(e)}$ there is one bifurcation point, $B_T^6$ at high eccentricity value.

\begin{figure}[tb]
\centering
\includegraphics[width=12cm]{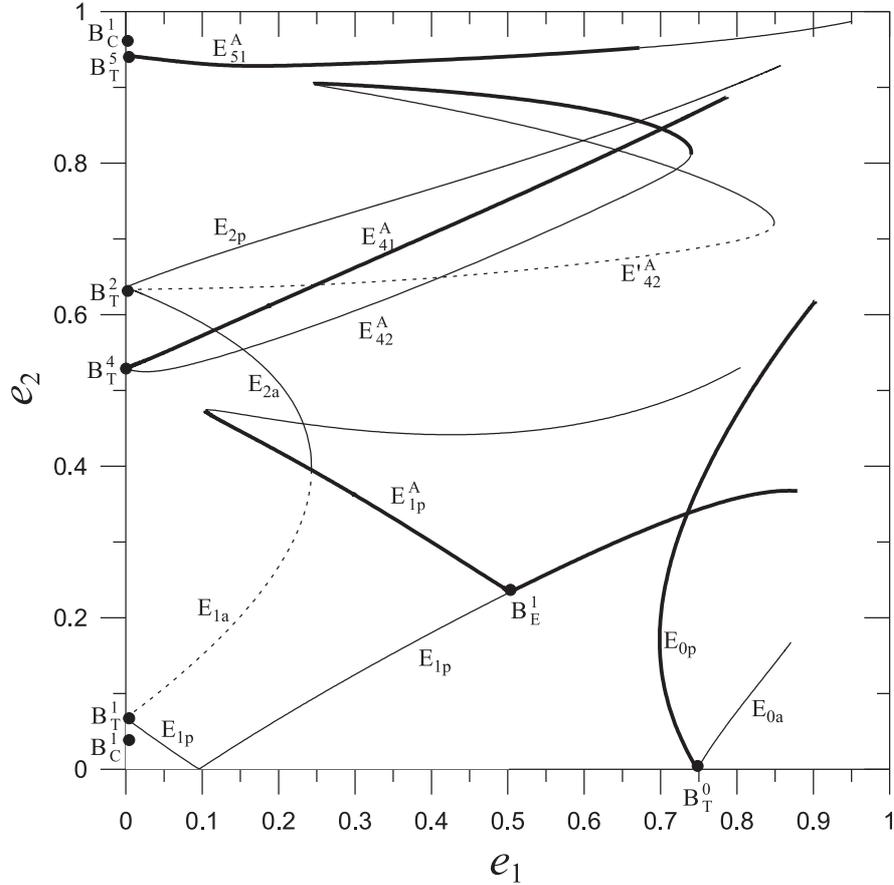}
\caption{Families of  2/1 resonant periodic orbits in the elliptic restricted problem. The critical orbits $B_T^i$, $i=0,1,2,4,5$ of the circular problem are indicated. Bold, thin and dotted face curves indicate stable, unstable and doubly unstable orbits respectively.} \label{FERPO}
\end{figure}

Let us consider a critical periodic orbit  $B_T^i$ of the circular
problem. The continuation of a family of periodic orbits of the
elliptic model, which bifurcates from a point $B_T^i$, is obtained
by increasing the eccentricity of Jupiter, starting from the zero
value and keeping its semimajor axis equal to unity, $a'=1$, so that
$n'=1$. This is a family of resonant periodic orbits, along which
{\it the eccentricity of Jupiter increases}. There are two
possibilities (Hadjidemetriou, 1993): the elliptic orbit of Jupiter
corresponds (i) to perihelion at $t=0$ or (ii) to aphelion. So,
there are two families of resonant periodic orbits of the elliptic
model which bifurcate from each point $B_T^i$. In one family,
denoted by $E_p$, Jupiter is at perihelion and in the other family,
denoted by $E_a$, Jupiter is at aphelion. Additionally, as we will
show in the following, families of asymmetric periodic orbits also
exist for the elliptic problem and can be generated in three
different ways. We will denote the asymmetric orbits by $E^A$.

The resonant families of the elliptic model will be presented in the space of the eccentricities of Jupiter and the small body. There are two bodies involved (Jupiter and the small body) and in all the following we will call $e_1$ the eccentricity of the inner of these bodies and $e_2$ the eccentricity of the outer body, irrespectively of whether the corresponding body is Jupiter or the massless body. We use this notation in order to have a direct comparison with the families of the general TBP presented in the following sections.

Let us start from the family $II_{(i)}$ of the circular model, and present the family of the elliptic model that bifurcates from the orbit $B_T^0$ (Figure \ref {FCRPO}a), in the space $e_1$-$e_2$. Now $e_1$ denotes the eccentricity of the small body (inner body) and $e_2$ denotes the eccentricity of Jupiter (outer body). The Family $II_{(i)}$ of the 2/1 resonant family of the {\it circular} model is located in Figure \ref{FERPO} on the axis $e_2=0$ (circular orbit of Jupiter). On this axis we present the point $B_T^0$ and show the two families of the {\it elliptic} model that bifurcate from this point. As we mentioned before, there are two families, $E_{0a}$ and $E_{0p}$, corresponding to perihelion and aphelion of Jupiter, respectively. Both families start from the eccentricity $e_1=0.75$ of the small body and the eccentricity $e_2$ of Jupiter starts from zero and increases along the family.

We come now to the families of the elliptic model that bifurcate from the 1/2 external resonant family $II_{(e)}$ of the {\it circular} model (Figure \ref {FCRPO}b). In this case, $e_1$ is the eccentricity of Jupiter (inner body) and $e_2$ is the eccentricity of the small body (outer body). The family $II_{(e)}$ is located in the Figure \ref{FERPO} on the axis $e_1=0$. Let us consider first the bifurcation from the symmetric families of the circular problem. We study the two critical points $B_T^1$ and $B_T^2$ (the point $B_T^3$ corresponds to very high eccentricities and we do not study it here). From the point $B_T^1$ there bifurcate two resonant families of symmetric periodic orbits of the elliptic model, the family $E_{1p}$, which starts having stable orbits, and the family $E_{1a}$ where the orbits are doubly unstable. From the critical point $B_T^2$ there also bifurcate two families of resonant symmetric periodic orbits of the elliptic model, the families $E_{2a}$ and $E_{2p}$; both are unstable.  It turns out that the family $E_{1a}$ that bifurcates from the point $B_T^1$ and the family $E_{2a}$ that bifurcates from the point $B_T^2$ coincide and form a single family that starts from $B_T^1$ and ends to $B_T^2$.

We come next to the asymmetric families of the elliptic model. There
exist two critical points on the asymmetric family $A$ in
Fig.\ref{FCRPO}, the points $B_T^4$ and $B_T^5$. As in the previous
cases, from each one of these points we have a bifurcation of two
resonant families of the elliptic model, which are asymmetric. From
the point $B_T^4$ we have the bifurcation of the family $E_{41}^A$
(stable) and the family $E_{42}^A$ (unstable). This latter family
has a complicated form, its stability changes three times and ends
at the point $B_T^2$. So it seems that from this latter point there
bifurcate, in addition to the two symmetric families, one more
asymmetric family. From the asymmetric point $B_T^5$ there bifurcate
two asymmetric families, $E_{51}^A$ and $E_{52}^A$. The family
$E_{51}^A$ is stable but family $E_{52}^A$ cannot be numerically
continued due its strong instability and, thus, is not presented in
Fig.\ref {FCRPO}.

Along the resonant families of the elliptic model it may also happen
to exist bifurcation points. Such a case is with the family
$E_{1p}$, where the critical point $B_E^1$ appears. This happens
because the stability type changes at this point. From this point we
have a bifurcation of the family $A_{1p}^A$ of asymmetric periodic
orbits. Voyatzis and Kotoulas (2005) showed that many families of
the external resonances have such critical points and conjectured
the bifurcation of asymmetric orbits.

\section{From the restricted to the general problem} \label{SectionCon}

\subsection{The rotating frame and periodicity conditions}

Let us consider three bodies with non zero masses, with one of them much more massive than the other two.  The more massive body with mass $m_0$ is the Sun, $S$, and the two small bodies, $P_1$ with mass $m_1$ and $P_2$ with mass $m_2$, will be called {\em planets}. In the following, $P_1$ will be the inner planet and $P_2$ the outer planet. If $m_1\neq 0$ and $m_2=0$ we have the restricted model where $P_1$ is the corresponding Jupiter and $P_2$ the small body, which moves initially in an outer orbit. If the Keplerian orbit of the bodies $S$ and $P_1$ is circular, we have the circular restricted problem and if it is elliptic, we have the elliptic restricted model. If $m_1=0$ and $m_2\neq 0$ then $P_2$ plays the role of Jupiter and the massless body $P_1$ evolves initially in an inner orbit.

In order to study the continuation of the periodic orbits from the restricted to the general problem we define a rotating frame of reference $xOy$, whose $x$-axis is the line $S-P_1$, with the center of mass of these two bodies at the origin $O$ and the $y$-axis is in the orbital plane of the three bodies. In this rotating frame the body $P_1$ is always on the $x$-axis and $P_2$ moves in the $xOy$ plane. We have four degrees of freedom and we use as coordinates the position $x_1$ of $P_1$, the coordinates $x_2$, $y_2$ of $P_2$ and the angle $\theta$ between the $x$ axis and a fixed direction in the inertial frame.

The Lagrangian in the above coordinates is (Hadjidemetriou 1975)
\begin{equation}
{\cal L}=\frac{1}{2}(m_1+m_2) \left [\frac{m_1}{m_0}(1-\mu)^2(\dot
r^2+r^2\dot\theta^2)+\frac{m_2}{m} \Big( \dot x_2^2+\dot
y_2^2+\dot\theta^2 (x^2+y^2)+2\dot\theta (x\dot y-\dot x y) \Big)
\right ]+ V,
\end{equation}
where $m=m_0+m_1+m_2$ is the total mass of the system, $\mu=m_1/(m_0+m_1)$, $r=x_1/\mu$ is the distance between the Sun and $P_1$ and
$$
V=-\frac{G m_0 m_1}{r}-\frac{G m_1 m_2}{\sqrt{((1-\mu)r-x_2)^2+y_2^2}}-\frac{G m_0 m_2}{\sqrt{(\mu r+x_2)^2+y_2^2}}.
$$
Note, that $\theta$ is ignorable and consequently the angular momentum $L=\partial {\cal L}/\partial\dot\theta$ is constant. We can use this angular momentum to reduce the number of degrees of freedom from four to three. In the reduced Lagrangian, called the {\it Ruthian}, the angular momentum constant $L$ is a fixed parameter. For this reason we take as normalizing conditions to fix the units of mass, length and time the conditions
\[
m=1,\:\:\:G=1,\:\:\:L=\mbox{constant}.
\]

Periodic orbits of period $T$ exist in the rotating frame and their initial conditions must satisfy the  periodic conditions (Voyatzis and Hadjidemetriou, 2005):
\begin{eqnarray}
\dot{x}_1(T)=\dot{x}_1(0)=0 ,& \;\;x_1(T)=x_1(0) , \nonumber \\
x_2(T)=x_2(0) ,&\;\;y_2(T)=y_2(0), \nonumber\\
\dot{x}_2(T)=\dot{x}_2(0) ,&\;\;\dot{y}_2(T)=\dot{y}_2(0). \nonumber
\end{eqnarray}
Thus, a periodic orbit is represented by a point in the 5-dimensional space
\begin{equation}
\Pi=\{(x_1(0),x_2(0),y_2(0),\dot{x}_2(0),\dot{y}_2(0))\}.
\end{equation}
When a periodic orbits is symmetric, i.e. it is invariant under the fundamental symmetry $\Sigma: (t,x,y) \rightarrow (-t,x,-y)$ (H\'enon, 1997; Voyatzis and Hadjidemetriou, 2005), we can always take as initial conditions $y_2(0)=0$ and $\dot x_2(0)=0$ and the dimension of the space of initial conditions $\Pi$ is reduced to three.  In the planetary problem, the orbits of $P_1$ and $P_2$ are almost Keplerian and we can present the families of periodic orbits in the projection plane of planetary eccentricities $e_1 - e_2$, which correspond to the initial conditions.

The monodromy matrix of the variational equations has now three pairs of eigenvalues ($\lambda_1$,$\lambda_2$), ($\lambda_3$, $\lambda_4$) and ($\lambda_5$, $\lambda_6$). Due to the existence of the energy integral it is $\lambda_5=\lambda_6=1$ and the stability of the periodic orbits is defined by the first two pairs as in the elliptic restricted problem (see section \ref{ERTBP}).

%%The linear stability of  of a periodic orbit can be determined through the computation of the stability indices $b_1$, $b_2$  (Hadjidemetriou, 2006). %%A periodic orbit is considered linearly stable if $$|b_i|<2, \;\;\forall i=1,2.$$ If one of the indices $b_i$ does not satisfy the stability %%condition, the periodic orbit is {\em simply unstable}. If both indices do not satisfy the stability condition the periodic orbit is {\em doubly %%unstable}.

\subsection{The continuation from the restricted to the general problem}

In general, the periodic orbits of the restricted problem are continued to the general problem. Particularly, it is proved by Hadjidemetriou (1975) that a periodic orbit of period $T$ of the circular restricted problem is continued to the general problem with the same period, by increasing the mass (e.g. $m_2$) of the initially massless body. The continuation is not possible only in the case where the period is a multiple of $2\pi$ (the period of the primaries). Such continuation forms monoparametric families of periodic orbits with parameter the mass of the small body ($m_2$), provided that the masses of the other two bodies are fixed. If we keep all masses fixed (and non zero), we obtain a monoparametric family of periodic orbits of the general problem, of the planetary type, along which the elements of the two planetary orbits vary. This family is represented by a smooth curve in the space $\Pi$ of initial conditions. The periodic orbits of the elliptic restricted problem are also continued to the general problem (Hadjidemetriou and Christides, 1975).

\begin{figure}[tb]
\centering
\includegraphics[width=10cm]{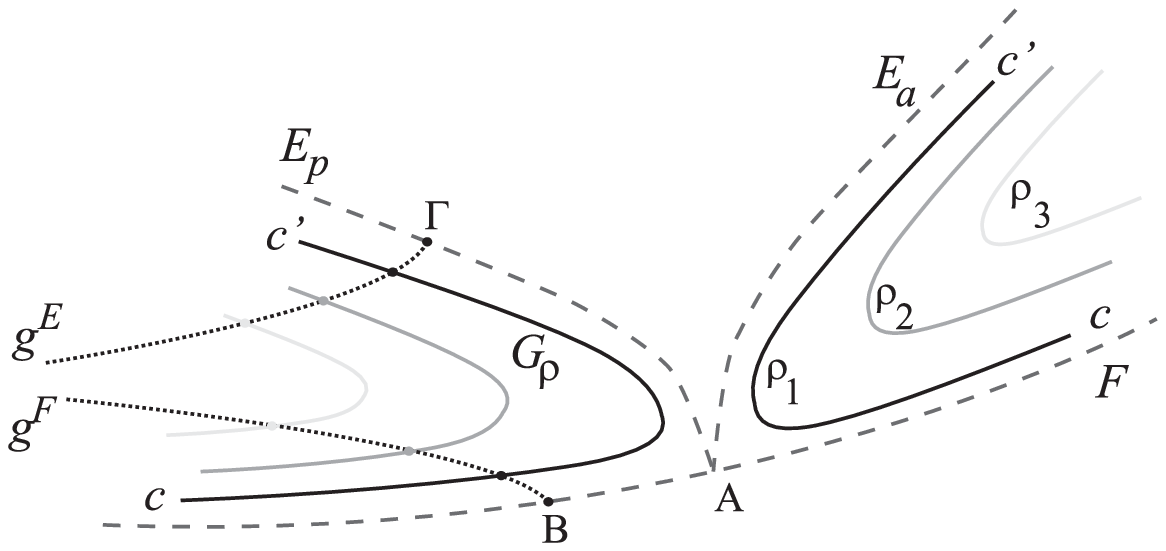}
\caption{The schematic generation of families of periodic orbits in the general problem from the restricted problem. Dashed curves indicate the families of the restricted problem and the solid curves indicates the families continued in the general problem (see text for detailed description).} \label{ConScene}
\end{figure}

The evolution of the characteristic curves of the families, as $m_2$ increases, is studied by Bozis and Hadjidemetriou (1976). In the present paper we study the continuation using a slightly different approach.  Provided that the planetary masses are small with respect to the mass of Sun, i.e. $m_1\ll m_0$ and $m_2\ll m_0$, it is found that the families of resonant periodic orbits in the space $\Pi$ depends on the ratio $\rho=m_2/m_1$ of the planetary masses rather than by their actual values (Beauge et al., 2003). Thus, we can use $\rho$ as a continuation parameter for a monoparametric family of periodic orbits.  If we add one more dimension to the space of initial conditions $\Pi$ in order to assign the value of $\rho$, we obtain an extended space $\Pi'$. In this extended space we can form characteristic surfaces of two-parametric families. In the following we will present families of periodic orbits considering sections of $\Pi'$ defined by fixing $\rho$ to a constant value.

\begin{figure}[tb]
\centering
\includegraphics[width=7cm]{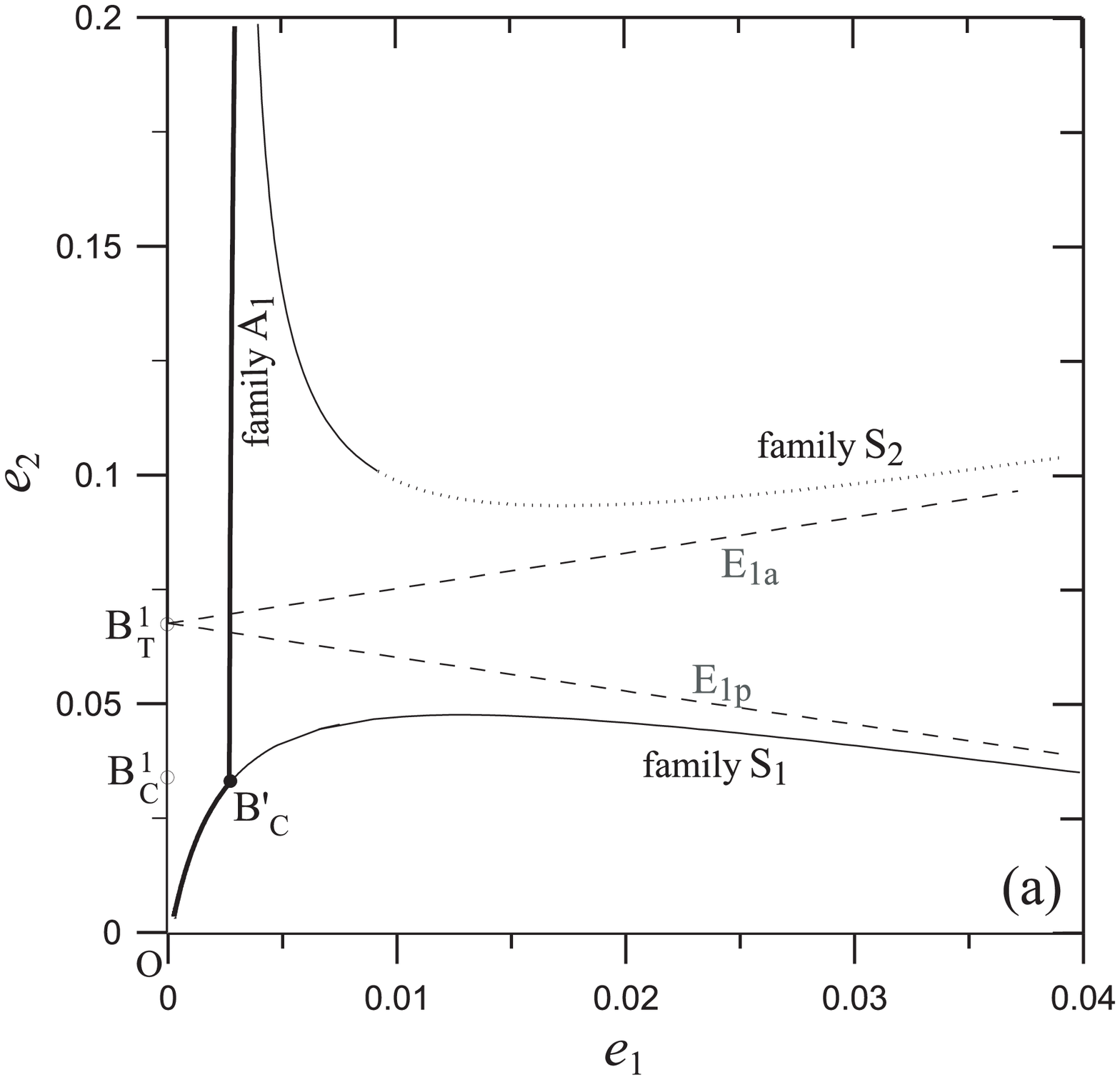} \hspace{1cm}
\includegraphics[width=7cm]{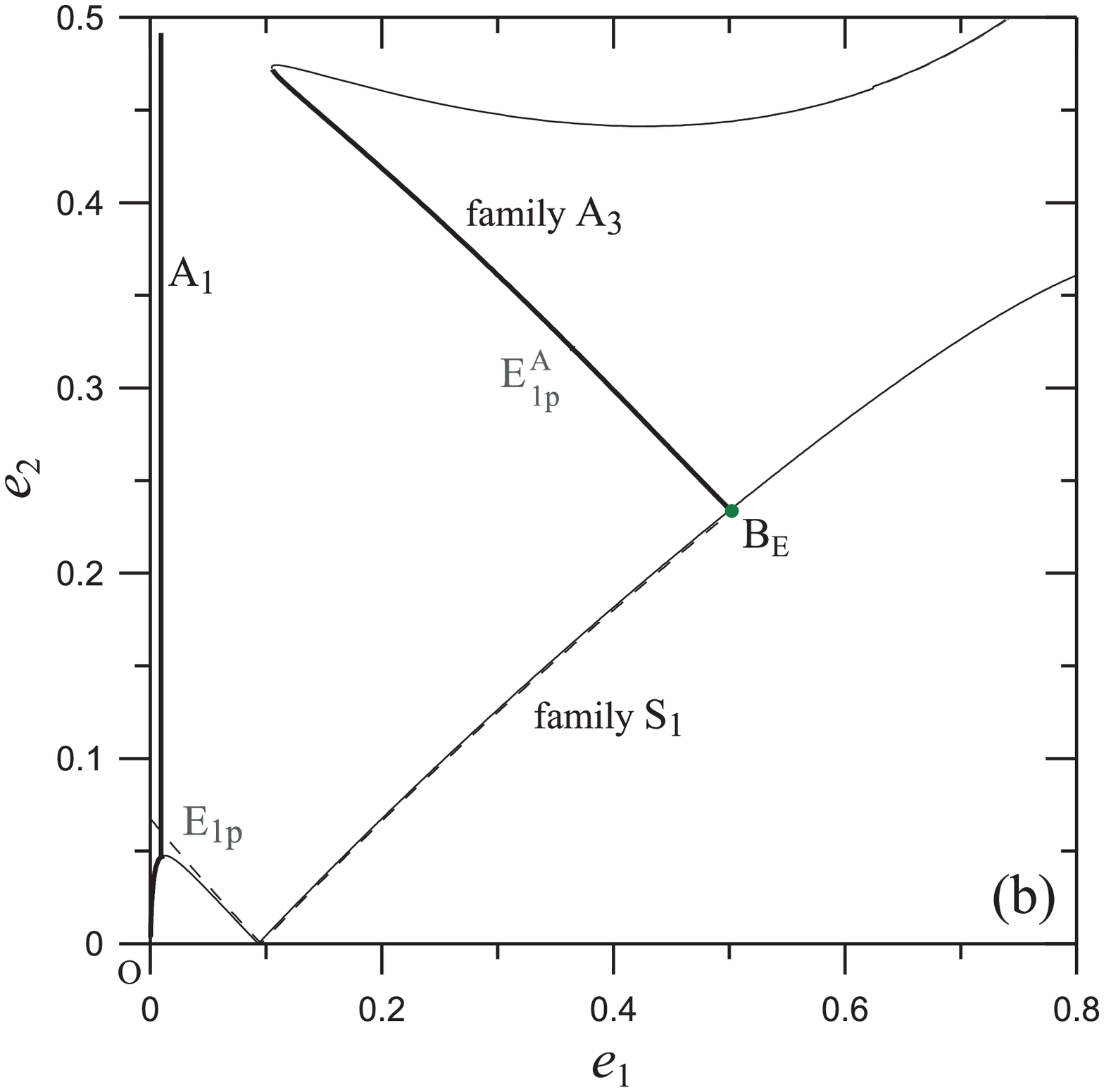} \\
\includegraphics[width=7cm]{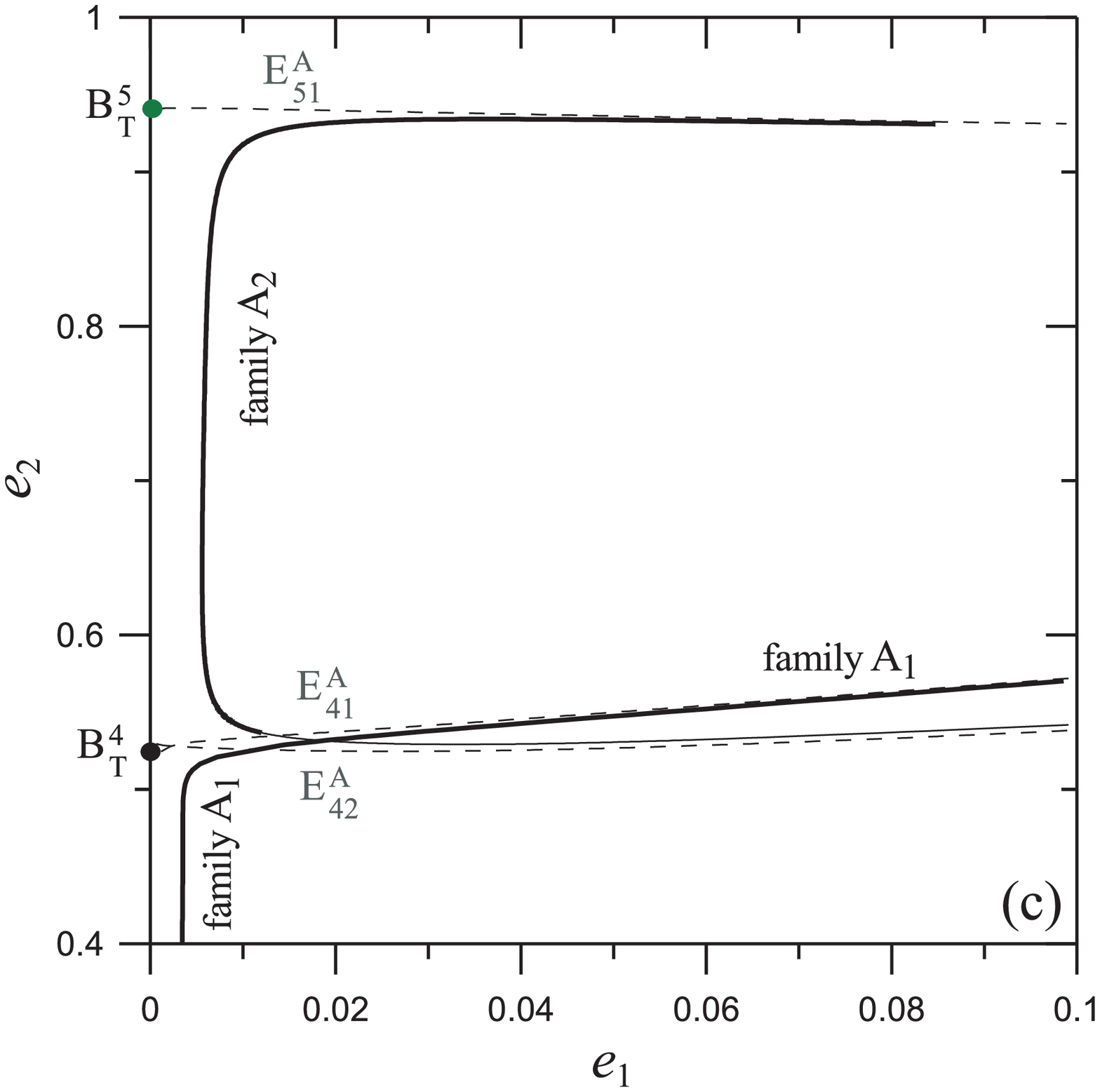} \hspace{1cm}
\includegraphics[width=7cm]{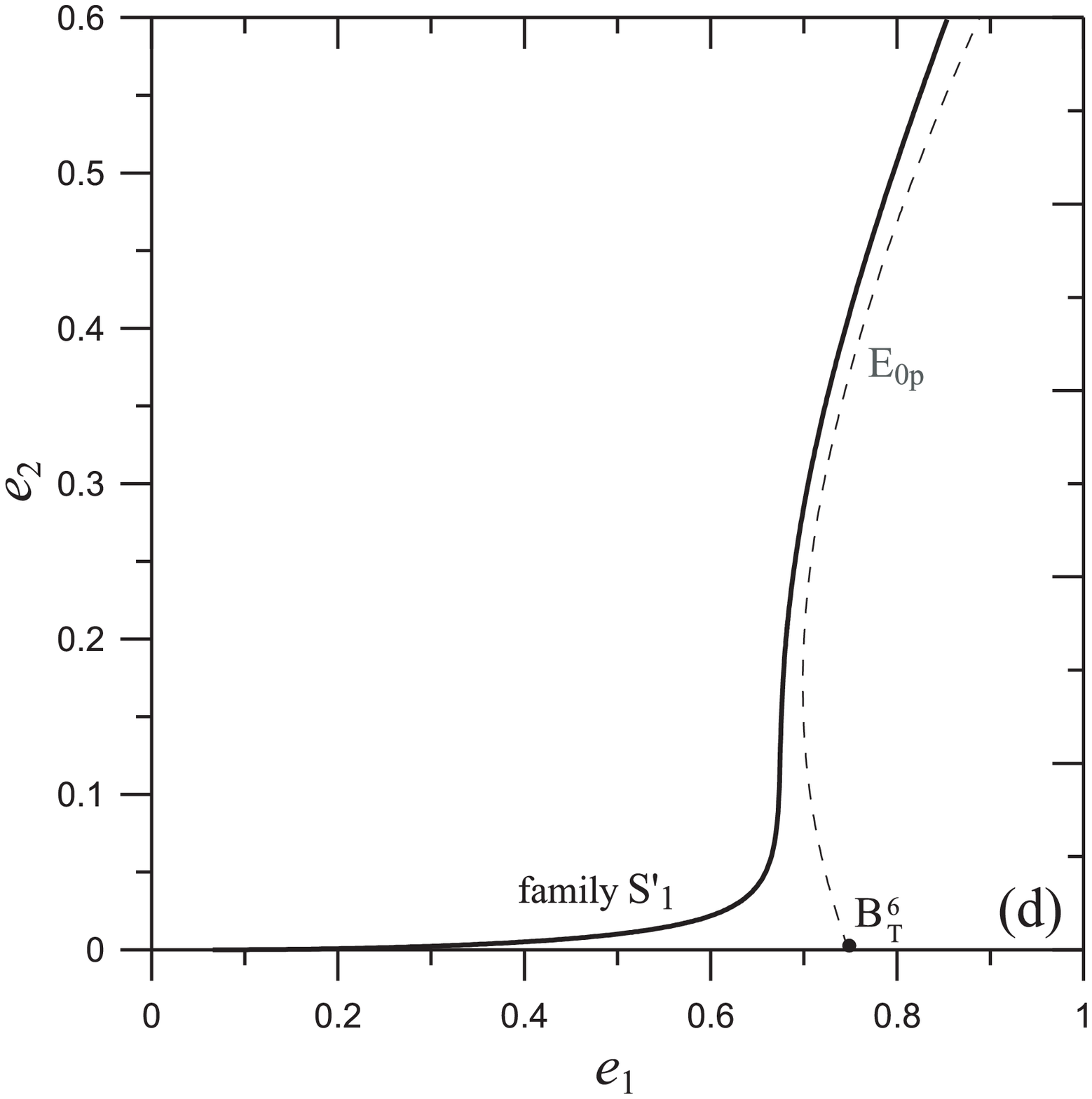}
\caption{Families of periodic orbits of the general problem (solid
curves) for $\rho=0.01$ in cases (a)-(c) (external resonance) and $\rho=100$ in case (d) (internal resonance). The families of the circular restricted problem lie along the axis $e_1=0$ and $e_2=0$ for the external and internal resonance, respectively. The families of the elliptic restricted problem are indicated by the dashed curves.} \label{FRG}
\end{figure}

In Fig. \ref{ConScene} it is shown schematically the continuation of
the families $F$ and $E$ of the circular and the elliptic restricted
problem, respectively, with e.g. $\rho=0$. The point A of the family
$F$ of periodic orbits of the circular model corresponds to an orbit
with period $T=2k\pi$, where $k$ is an integer. At this point the
families $E_p$ and $E_a$ bifurcate to the elliptic restricted
problem. By giving mass to the small body, any orbit of $F$ with
$T\neq 2k\pi$, e.g. the orbit B, is continued to the general problem
with parameter the mass ratio $\rho$ and the monoparametric family
$g^F$ is formed. A family $g^F$ exist for any initial orbit of $F$,
except for the orbit A. If we start now from a point on the family
$g^F$ and keep $\rho$ fixed, we obtain a monoparametric family $c$.
The same continuation holds for the periodic orbits of the families
$E_p$ and $E_a$ of the elliptic restricted problem and for a fixed
$\rho\neq 0$ we get the characteristic curve $c'$. The individual
parts of $c$ and $c'$ join smoothly to each other forming the
families $G_\rho$ for the general problem and for any fixed mass
ratio $\rho\ll 1$. The periodic orbit A is a {\em singularity} for
the continuation resulting in the formation of a gap between the
left and right characteristic curves $cc'$.

\subsection{Families of 2/1 and 1/2 orbits continued from the restricted problem to the general problem}
Concerning the continuation of the resonant families, which are presented in sections \ref{CRTBP} and \ref{ERTBP}, to the general problem, singularities are expected at the bifurcation points $B_T$ (see Fig. \ref{FERPO}) for both internal (2/1) and external (1/2) resonances, where is $T=2k\pi$. In the following, we study four different cases of continuation presented in Fig. \ref{FRG}. We remind that the families of the circular restricted problem are continued as curves close to the vertical and the horizontal axes in the plane $e_1-e_2$, for the external ($e_1=0$) and the internal resonances ($e_2=0$), respectively. In our computations we vary $\rho$ by changing the value of one of the planetary masses and such that $\max(m_1,m_2)=10^{-3}$.

In the panel (a) of Fig. \ref{FRG} we show the continuation close to the bifurcation point $B_T^1$, which belongs to the symmetric family $II_{(e)}$ of the circular problem and is the starting point for the families $E_{1p}$ and $E_{1a}$ of the elliptic problem. By giving mass to the small body (i.e. $\rho\neq 0$) the part ($O B_T^1$) of the family $II_{(e)}$ and the family $E_{1p}$ are continued and they join smoothly forming the symmetric family $S_1$.  The change of stability that is shown in family $II_{(e)}$ at the point $B_C^1$ is also obtained in the family $S_1$. Namely, the bifurcation point $B_C^1$ (periodic orbit of critical stability) continues as a bifurcation point $B'_C$ in the general problem too. The asymmetric family $A$ of the circular restricted problem, which bifurcates from $B_C^1$, is continued smoothly to the general problem as an asymmetric family which bifurcates from the $B'_C$. Similarly to the formation of $S_1$, the unstable part of the family $II_{(e)}$ above the point $B_T^1$ and the family $E_{1a}$, which is doubly unstable, are continued and form the family $S_2$. Note that the different stability types of the two families results to a periodic orbit of critical stability in the family $S_2$.

In family $E_{1p}$ there is a change in stability at a point $B_E$, where the asymmetric family $E_{1p}^A$ bifurcates for the elliptic problem. As $\rho$ takes a positive value, $B_E$ continues as a critical point in the general problem and the asymmetric family $A_3$ bifurcates from it (Fig.  \ref{FRG}b). Actually, $A_3$ can be considered as the continuation of the family $E_{1p}^A$. No singularities are obtained in this case. Such a type of continuation explains the existence of the asymmetric planetary corotations found by Michtchenko et al. (2006) at the 3/2 resonance, which in the elliptic restricted problem shows an asymmetric family similar to $E_{1p}^A$. We note that asymmetric periodic orbits were known only for resonances $p/q$ with $q=1$.

As it is mentioned above, the asymmetric family $A$ continues smoothly to the general problem (as family $A_1$) starting from its bifurcation point $B_C^1$ (see Fig. \ref{FRG}a) . However, the family $A$ contains the critical point $B_T^4$ (see Fig. \ref{FCRPO}), which admits a singularity for the continuation to the general problem. In the panel (c) of Fig. \ref{FRG}, the gap formed in the above singularity is shown. The asymmetric family $E_{41}^A$, which bifurcates from $B_T^4$, continues and completes the asymmetric family $A_1$ of the general problem. Additionally, the part of family $A$ located between the bifurcation points $B_T^4$ and $B_T^5$, the family $E_{42}^A$ and the family $E_{51}^A$ continue in the general problem, join smoothly and form the asymmetric family $A_2$. Again we obtain a change of stability along $A_2$ due to the different types of stability on the families of the restricted problem (family $A$ is stable while family $E_{42}^A$ is unstable).

In the case of the internal resonance, the continuation to the general problem shows the same characteristics as in the external resonances. We remark that in this case only symmetric periodic orbits appear. In Fig. \ref{FRG}d we show the formation of the symmetric family $S'_1$ of the general problem, after the continuation of the families $II_{(i)}$ and $E_{0p}$ of the restricted problem. We note that since $\rho=\infty$ in this case, the continuation maybe assumed as varying $1/\rho$.

\section{Continuation within the framework of the general problem} \label{GTBPbif}

In the previous section we considered the continuation of families of periodic orbits when we pass from the restricted to the general problem. The situation, which was described, is referred to small values of the mass of one of the planets, namely $\rho\ll 1$ or $\rho\gg 1$. Now we consider the case of the external resonance and its families generated for $\rho>0$. As the parameter $\rho$ increases, the families evolve and new bifurcations and structure changes are possible. We follow the evolution of the 2/1 resonant families, which are presented in the previous section, and we show in the following how the new structures are formed by increasing $\rho$. We remind that as $\rho \rightarrow \infty$ we approach the internal resonance of the restricted problem.

\begin{figure}[tb]
\centering
\includegraphics[width=16cm]{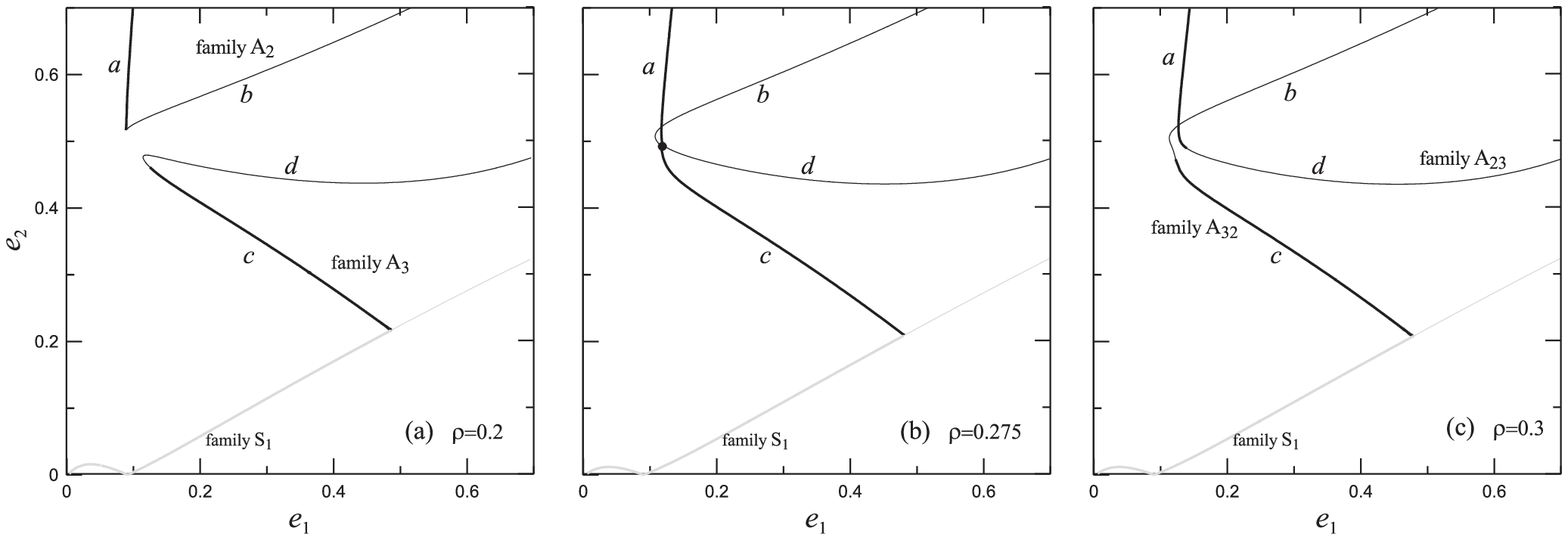}
\caption{The ``collision--bifurcation'' of the families $A_2$ and $A_3$ at $\rho=\bar\rho_2$ and the formation of the new families $A_{23}$ and $A_{32}$. a) $\rho=0.2<\bar\rho_1$ b) $\rho=0.275\approx\bar\rho_1$ and c) $\rho=0.3>\bar\rho_1$.} \label{FGG1}
\end{figure}

\begin{figure}[tb]
\centering
\includegraphics[width=16cm]{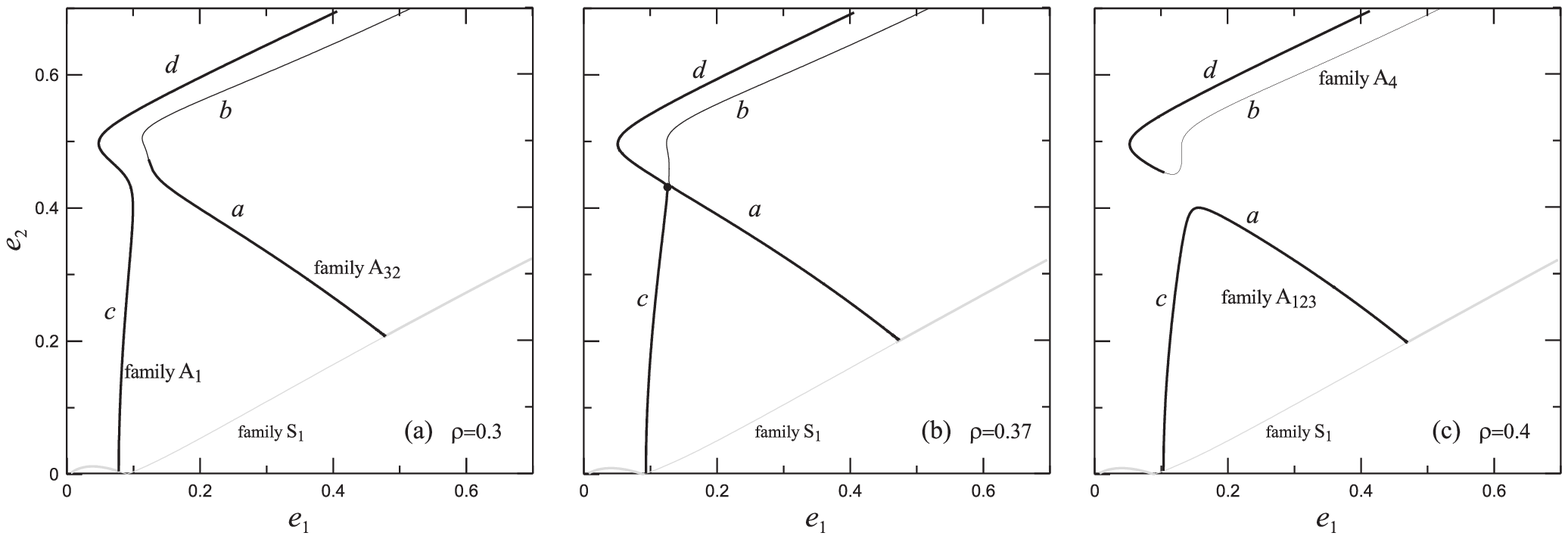}
\caption{The ``collision--bifurcation'' of the families $A_1$ and $A_{32}$ at $\rho=\bar\rho_2$ and the formation of the new families $A_4$ and $A_{123}$. a) $\rho=0.3<\bar\rho_2$ b) $\rho=0.37\approx\bar\rho_2$ and c) $\rho=0.4>\bar\rho_2$.} \label{FGG2}
\end{figure}

Figure \ref{FGG1}a shows the asymmetric families $A_2$ and $A_3$ for
$\rho=0.2$. Up to this value not any structural change of the
families occur. In both families the stability changes and,
consequently, we distinguish in each family two parts, parts $a$ and
$b$ for $A_2$ and $c$ and $d$ for $A_3$. For the critical value
$\rho=\bar\rho_1\approx 0.275$ the two families collide at a point
in the space of initial conditions $\Pi$ (Fig. \ref{FGG1}b). This
point corresponds to the orbit of critical stability and is the
intersection point of the parts $a$--$d$. For $\rho\gg \bar\rho_1$
(see Fig. \ref{FGG1}c) the part $a$ of family $A_2$ and the part $d$
of family $A_3$ join together and form the family $A_{23}$.
Similarly, the parts $c$ and $b$ of the families $A_3$ and $A_2$,
respectively, form the family $A_{32}$. Note that the new families
are separate and their intersection in the plane $e_1-e_2$ is due to
the projection.

The evolution of the family $A_1$, as $\rho$ increases, take place smoothly without structural changes up to $\rho=\bar\rho_2\approx 0.37$. In Figure \ref{FGG2}a, which corresponds to $\rho=0.3$, it is shown that the family $A_1$ has come close to the family $A_{32}$, which is generated after the bifurcation at $\rho=\bar\rho_1$. At $\rho=\bar\rho_2$ the two families collide and, similarly to the previous case, for $\rho>\bar\rho_2$ we obtain two new families, the family $A_4$ and the family $A_{123}$ (Fig. \ref{FGG2}c). In this case only the family $A_{32}$ has an orbit of critical stability,
which separates the family in a stable ($a$) and in an unstable part ($b$). The family $A_1$ is whole stable and there is no clear border between its parts $c$ and $d$. After the bifurcation, an orbit of critical stability is shown only in the new family $A_4$, while the family $A_{123}$ is whole stable and starts and ends at bifurcation points of the symmetric family $S_1$. Note that these bifurcation points originate to the bifurcation points $B_C^1$ and $B_E$ of the circular and the elliptic, respectively, restricted problem.

\begin{figure}[tb]
\centering
\includegraphics[height=7cm]{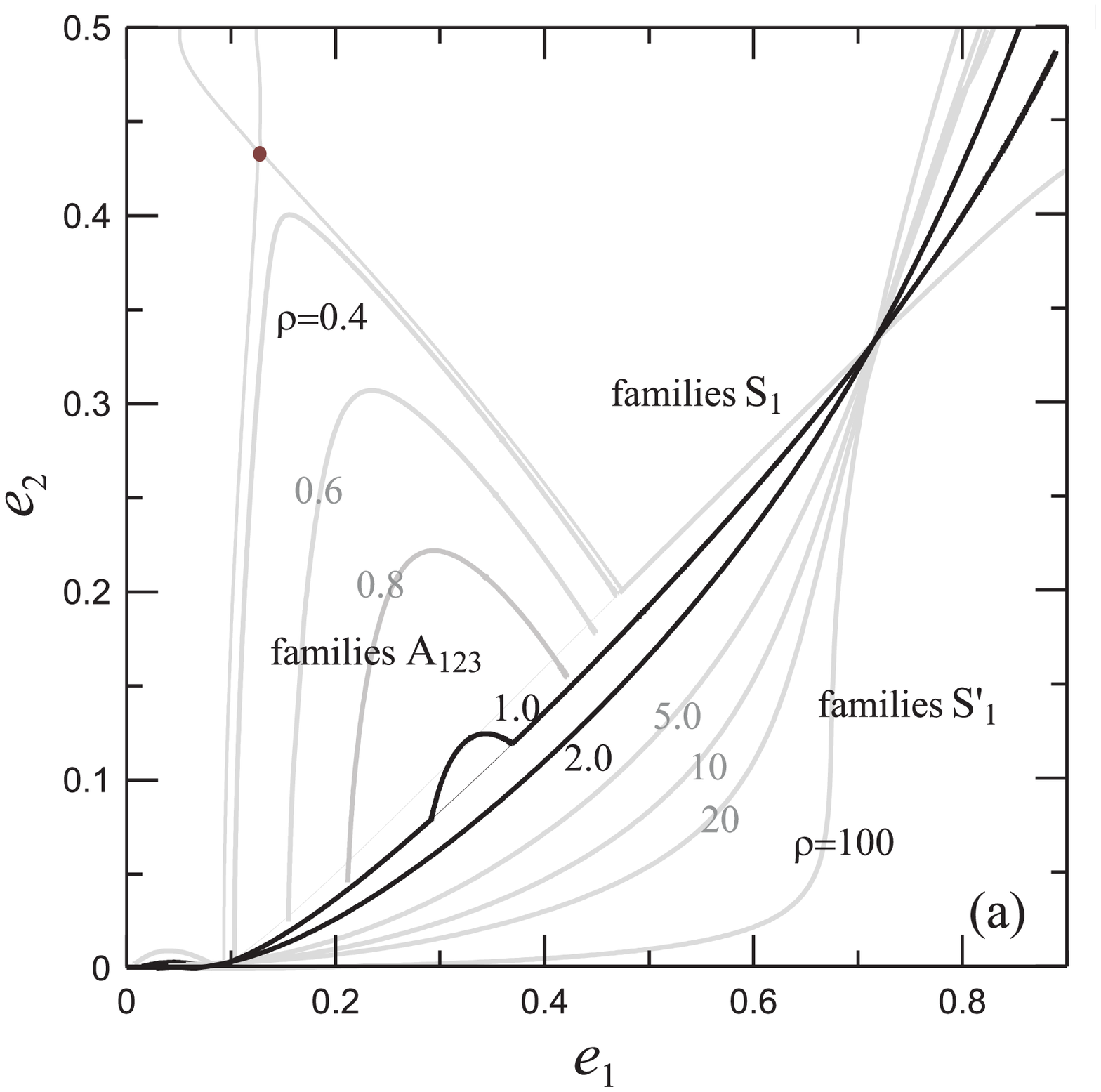} \hspace{1cm}
\includegraphics[height=7cm]{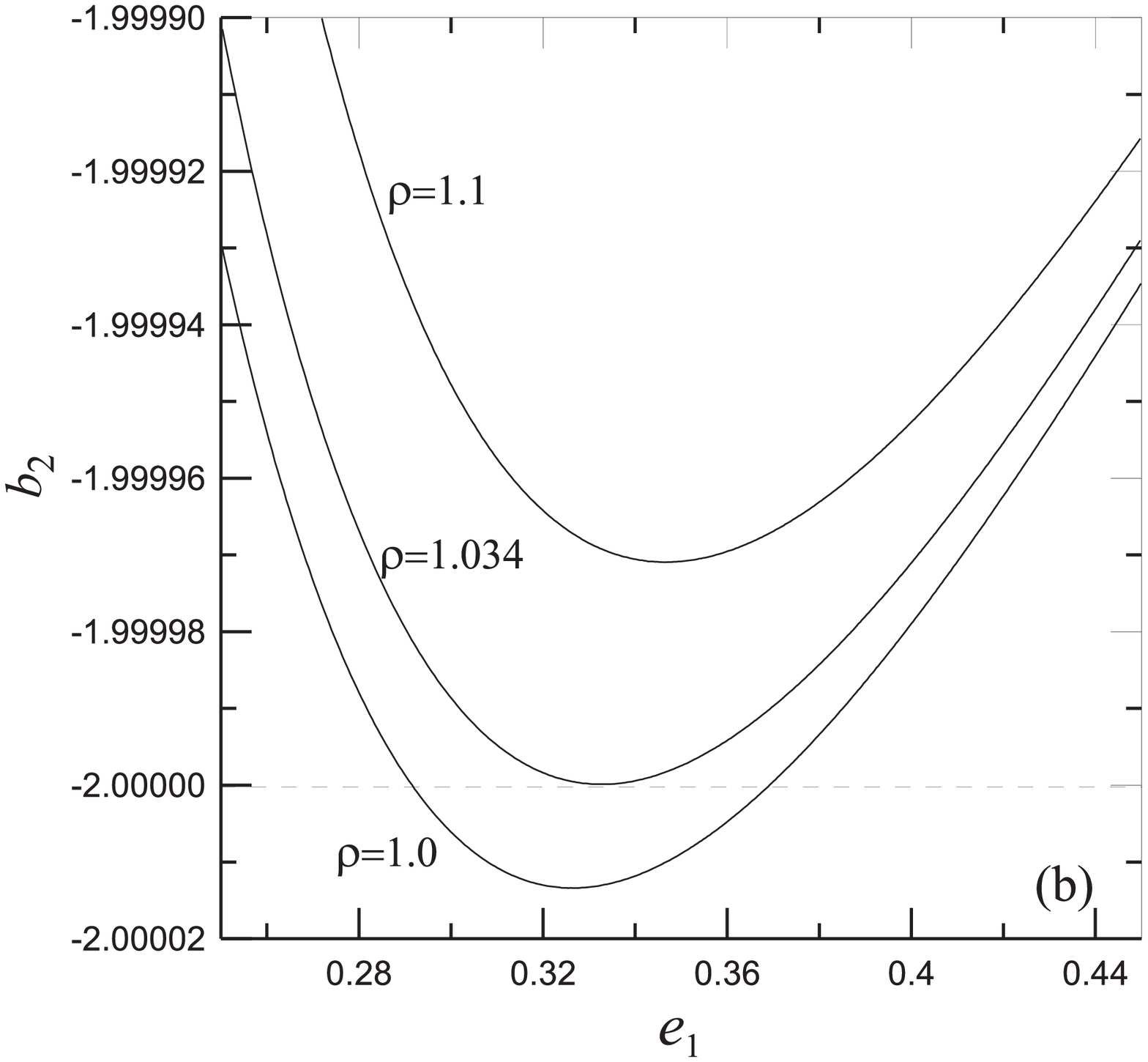} \\
\caption{a) The evolution of the families $A_{123}$ and $S_1$ as $\rho$ passes the critical value $\bar\rho_3=1.034$ and takes large values. b) The stability index $b_2$ along the family $S_1$, which determines the interval of instability ($b_2<-2$) and the bifurcation points (at $b_2=-2$) of the family $A_{123}$.} \label{FGG3}
\end{figure}

Now we restrict our study to the evolution of the family $A_{123}$
for $\rho>\bar\rho_2$. As it is shown in Fig. \ref{FGG3}a, as $\rho$
increases, the ending points of the family move on along the family
$S_1$ in opposite direction and the family shrinks and, finally
disappears at $\rho=\bar\rho_3\approx 1.034$. In Fig. \ref{FGG3}b we
present the above transition by considering the stability indices
$b_1, b_2$ for the orbits along the family $S_1$ (see
Hadjidemetriou, 2005). The horizontal axis indicates the
eccentricity of the periodic orbits along the family $S_1$ and the
value of the corresponding stability index $b_2$ is presented on the
vertical axis. For all orbits it holds $|b_1|<2$ and, thus, the
condition $|b_2|<2$ is sufficient and necessary for linear
stability. In Figure \ref{FGG3}b we obtain that unstable orbits
exist for $\rho<\bar\rho_3$ in a part of the family $S_1$ defined by
the $e_1$ interval where $b_2<-2$. As $\rho$ increases, the curve of
$b_2$ values is raised continually and for $\rho>\bar\rho_3$ is
located above the value $b_2=-2$. Therefore, the unstable part of
$S_1$ disappears and, consequently, the family $A_{123}$ disappears
too.

Actually for $\rho>\bar\rho_3$, the planet $P_1$ becomes the small body and we pass to the case of the internal resonance. The remaining family $S_1$ evolves smoothly, as $\rho$ increases, and should be assumed as a family $S'_1$ of the internal resonance, which approaches the families $II_{(i)}$ and $E_{0p}$ of the restricted problem as $\rho \rightarrow \infty$ (see sections \ref{CRTBP},\ref{ERTBP} and Fig. \ref{FRG}d).

The above described bifurcation scheme explains completely the origin and the structure of corotations found by Beauge et al. (2006). Following the other asymmetric families, which are coming from the restricted problem, we obtain new ``collision--bifurcations'' as $\rho$ increases and new structures of characteristic curves are formed. We have found e.g. that the families $A_4$ and $A_{23}$ collide for $\rho\approx 0.45$ and generate the miscellaneous family found by Voyatzis and Hadjidemetriou (2005) and called ``$A_2$'' there in.

\section{Discussion and Conclusions} \label{CD}
The method of continuation of periodic orbits has been applied in the present work in order to study the generation and the structure of families of periodic orbits in the general TBP, starting from the restricted problem. We considered the planar TBP of planetary type, referred to a rotating frame and studied resonant planetary motion. In particular, we studied the 2/1 resonant periodic orbits, but the method we used can be applied to all other resonances. We considered planar motion only.

We started from the circular restricted TBP and computed the basic families of 2/1 resonant periodic orbits, both for the inner orbits (inside Jupiter) and the outer orbits (outside Jupiter). Along these families the resonance is almost constant, but the eccentricity of the small body increases and may take high values. The basic families are symmetric with respect to the rotating $x$-axis, but asymmetric families also exist in cases of resonances of the form $1/q$. On these families, we found the periodic orbits with period equal to $2\pi$ which, in the normalization we are using, are the bifurcation points to resonant 2/1 (or 1/2) families of periodic orbits of the elliptic restricted model, along which the eccentricity of Jupiter varies, starting from zero values. Two such families bifurcate from each of these critical points. In particular, we have the following cases:  a) an asymmetric periodic orbit of period $2\pi$ of the circular restricted problem is continued to the elliptic restricted problem as asymmetric also,  b) a symmetric periodic orbit of period $2\pi$ of the circular restricted problem can be continued to the elliptic problem as an asymmetric one - this exceptional case is verified only up to the accuracy of the numerical computations  and c) a symmetric periodic orbit of the elliptic restricted problem, which is of critical stability, can be continued to the elliptic problem as an asymmetric one. The last case indicates the existence of asymmetric periodic orbits in the elliptic problem and in particular to resonances which are not necessarily of the form $1/q$. In this way we obtain a clear picture of all the resonant families of the restricted problem, and in this way we can make a complete study of the resonant families of the general problem, by giving mass to the massless body of the restricted model.

The continuation of the families of periodic orbits of the restricted problem to the general problem follows the  scenario described in the paper of Bozis and Hadjidemetriou (1976).  Particularly, the families of the general problem originate from two families of the restricted problem, one of the circular problem and one of the elliptic problem. There is no essential difference in the continuation between symmetric and asymmetric periodic orbits. The asymmetric orbits, which bifurcate from symmetric orbits of the elliptic problem, are continued smoothly in the general problem. The stability of periodic orbits is preserved after the continuation from the restricted to the general problem. In the paper mentioned above, a case of continuation without the formation of a gap is described. We found this case only at bifurcation points of the families of the elliptic restricted problem (see Fig.\ref{FRG}b).

After the continuation of the periodic orbits of the restricted problem to the general problem, by giving a very small mass to the massless body, we studied how these families evolve when we increase the mass of the small body. Particularly we studied such an evolution by considering as a parameter the planetary mass ratio $\rho=m_2/m_1$, keeping $m_i\ll 1, i=1,2$. In our numerical computations we considered $\max(m_1,m_2)=10^{-3}$ and we varied $\rho$ by varying only the mass of the smaller body. Starting from $\rho=0$ (external resonances of the restricted problem) and increasing its value, we found that the characteristic curves of two different families can collide in the space of initial conditions. At these points a bifurcation takes place ({\em collision--bifurcation}) causing a topological change in the structure of the colliding families and the formation of new families. As $\rho \rightarrow \infty$ we approximate the families of the internal resonances of the restricted problem. Since asymmetric periodic orbits are not known for the internal resonances, it would be interesting for a future work a study of the evolution and the bifurcation of all asymmetric families up to values of $\rho>1$ and, furthermore, as $\rho \rightarrow \infty$.

In general, the continuation and the evolution the families of the external resonances ($\rho<1$) in the general TBP show more rich dynamics (number and structures of families and bifurcations) in comparison with the case of internal resonances. It is a fact that the complexity of the dynamics of continuation and evolution of families of periodic orbits depends on the existence of critical orbits of period $T=2k\pi$, orbits of critical stability and asymmetric periodic orbits. For example the family $S'_1$ of the 2/1 internal resonance evolves smoothly as $\rho$ changes and no any bifurcations are obtained in the domain $(\bar\rho_3,\infty)$. Also we can show that a similar smooth continuation and evolution holds for the family $I_{(e)}$ (parts $I_{(ea)}$ and $I_{(eb)}$ in Fig. \ref{FRG}). As $\rho$ starts from zero and tends to infinity, the family evolves smoothly and tends to the family $I_(i)$ (parts $I_{ia}$ and $I_{ia}$). No any bifurcations take place, while the collision orbit, which separates the parts (a) and (b) of the families, exists for all values of $\rho$. This family is equivalent to the family of ($\pi,\pi$)- corotations indicated by Beauge et al (2006) and has a similar evolution as that of the 3/1 family $S_2$ of ($\pi,0$)- corotations shown in Voyatzis (2008).

Apart from collision--bifurcations, structural changes in the characteristic curves of periodic orbits occur when their bifurcation points disappear. We showed how the asymmetric family $A_{123}$, which starts and ends at bifurcation points on the symmetric family $S_1$, shrinks and finally disappears after the collapse of both bifurcation points. An interesting case occurs when the starting and the ending points of a family are bifurcation points that belong to different families. In this case it is possible, for a critical value of $\rho$, one of the points to disappear or to become a bifurcation point for another family. Such a bifurcation has been observed in the 3/1 resonance (Voyatzis, 2008) and cause global changes in the structure of the characteristic curves.

Although we presented results that are associated with the 2/1 resonance, we can claim that similar features of continuation are present in other resonances. Our numerical study can not be extended efficiently up to high eccentricities ($e_1 \rightarrow 1$ or $e_2 \rightarrow 1$), due to computation restrictions. We believe that the features of the families in these regions can be revealed by considering  the rectilinear restricted problem beside the circular and the elliptic one.

\end{document}